\documentclass{article}
\usepackage{graphicx}
\usepackage{subfigure}
\usepackage{url}
\usepackage{float}
\usepackage{multirow}
\usepackage{array}
\usepackage{hyperref}

\makeatletter
    \newcommand\figcaption{\def\@captype{figure}\caption}
    \newcommand\tabcaption{\def\@captype{table}\caption}
\makeatother

\begin{document}

\title{On the Monotonicity of Work Function in k-Server Conjecture}
\author{Ming-Zhe Chen 
\\DTech Inc.
\\ 5751 Sells Mill Dr 
        \\ Dublin, OH 43017, USA
\\ thedtech@yahoo.com}
\maketitle

\begin{abstract}
This paper presents a mistake  in work fuction algorithm of k-server 
conjecture.  That is, the monotonicity of the work fuction is not always
true.
\end{abstract}

\section{Introduction}
The k-server conjecture has not been proved. 
A lot of literature deal with the k-server conjecture 
\cite{koutsoupias} \cite{manasse} \cite{larmore} \cite{borodim} 
and references therein.
In \cite{koutsoupias}, the work function algorithm (WFA) is so far 
the best determined algorithm for this problem.
In \cite{koutsoupias}, there are the facts as follows (excerpts from \cite{koutsoupias}):
\begin{quote}
Fact 3 For a work function $w$ and two configurations $X, Y$
\begin{equation}
 w(X) \leq w(Y) + D(X, Y) 
\end{equation}
Consider a work function $w$ and the resulting work function 
$w'$ after request $r$.
By Fact 3 we get
\begin{equation}
w' (X)= \min_{x \in X} \{w(X-x+r) + d(r,x)\}  \geq w(X)
\end{equation}
which translats to:  \\
Fact 4 Let $w$ be a work function and let $w'$ be the resulting work function
after request $r$.  Then for all configurations $X$ :
\begin{equation}
w' (X) \geq w(X) 
\end{equation} 
\end{quote}
But Fact 4 is not true. That is,  the monotonicity of work function is not true.

\section{The monotonicity of work function}
Fact 4 is not true because (2) is incorrect.  From
\begin{equation}
 w' (X)= \min_{x \in X} \{w'(X-x+r) + d(r,x)\}
\end{equation}
we can get (it is Fact 3, here it is true):
\begin{equation}
w' (X) \leq w'(X-x+r) + d(r,x)  
\end{equation}
That is (because $w'(X-x+r) = w(X-x+r)$):
\begin{equation}
w' (X) \leq w(X-x+r) + d(r,x) 
\end{equation}
Assume that $Y$ is a configuration which makes the minimum of $w'(X)$, so
\begin{equation}
 w' (X)= w(Y) + D(X, Y) 
\end{equation}
But we cannot get $w' (X) \geq w(X)$ (Fact 4) from (7) based on Fact 3
because Fact 3 is incorrect for this case. 
If Fact 3 is derived from 
$ w(X)= \min_{x \in X} \{w(X-x+r) + d(r,x)\} $, it is true. 
But it cannot be used universally in all other cases
because Fact 3 is derived under some conditions\\
It is known
\begin{equation}
w (X)= \min_{x \in X} \{w(X-x+r') + d(r',x)\} 
\end{equation}
where $r'$ is the request before the request $r$. 
Assume that $Z$ is a configuration which makes the minimum of $w(X)$, so
\begin{equation}
w (X)= w(Z) + D(X, Z) 
\end{equation}
In order for Fact 4 to be true, we have to prove the following:
\begin{equation}
w' (X)= w(Y) + D(X, Y)  \geq w(Z) + D(X, Z) 
\end{equation}
Unfortunately, the above is not always true.\\
We give a concrete counterexample as follows. \\
A 5-node weighted undirected graph. The node set is ${a,b,c,d,e}$.
The distances (edge weights) are as follows.
\[ d(a,b)=1, d(a,c)=7, d(a,d)=5,d(a,e)=8,d(b,c)=4, \]
\[d(b,d)=2, d(b,e)=10,d(c,d)=3, d(c,e)=9, d(d,e)=6 \]
Consider 3-servers on this graph. The initial configuration is $abc$ 
and the request sequence are
\[e,d,a,c,b,d\]
In the folloing table we give values of work functions 
corresponding to all  3-node 
configurations and all request sequence.
\newline
\newline
Table: Values of Work Functions for 3-servers\\
\newline
\begin{tabular}{c c c c c c c c c c c c}
\hline
\multicolumn{12}{c}{Configuration}\\ 
Request&i&$abc$&$abd$&$abe$&$acd$&$ace$&$ade$&$bcd$&$bce$&$bde$&$cde$\\
\hline
$\phi$ & 0&0&3&9&2&10&11&5&8&11&10 \\
$e$     &1 &16&15&9&16&10&11&14&8&11&10\\
$d$    &2&18&15&13&16&14&11&14&12&11&10\\
$a$    &3&18&15&13&16&14&11&17&15&12&18\\
$c$    &4&18&20&18&16&14&17&17&15&18&18\\
$b$   &5&18&20&18&18&16&19&17&15&18&17\\
$d$   &6&20&20&21&18&22&19&17&19&18&17\\
\hline
\end{tabular} 
\newline
\newline
From the above table we can see
\[w_{edacb}(cde) < w_{edac}(cde) \]
so the Fact 4 is not always true. That is, 
the work function does not have the monotonicity.
In paper \cite{koutsoupias}, all theorems which are proved based
on the monotonicity have to be re-examined.
In paper \cite{koutsoupias}, the extended cost may overestimate the online
cost. We still think WFA would be $k$-competitive.
\bibliography{kserver}

\begin{thebibliography}{1}

\bibitem{borodim}
Allan Borodim and Ran El-Yaniv.
\newblock {\em Online Computation and Competitive Analysis}.
\newblock Cambridge University Press, 1998.

\bibitem{manasse}
Mark S.~Manasse et~al.
\newblock Competitive algorithms for server problem.
\newblock {\em Journal of Algorithms}, 11:208--230, 1990.

\bibitem{koutsoupias}
Elias Koutsoupias and Christos Papadimitriou.
\newblock On the k-server conjecture.
\newblock {\em Journal ACM}, 42:971--983, 1995.

\bibitem{larmore}
Lawrence~L. Larmore and Lames~A. Oravec.
\newblock T-theory applications to online algorithms for the server problem.
\newblock {\em arXiv:cs/0611088v1 [cs.DS] 18 Nov}, 2006.

\end{thebibliography}
\bibliographystyle{plain}
\end{document}